\documentstyle[preprint,prl,aps]{revtex}

\begin{document}
\preprint{\vbox{\hbox{November 1995}\hbox{IFP-714-UNC}\hbox{VAND-TH-95-5}}}
\draft
\title{Constraining $\alpha_s(M_Z)$ from the Hidden Sector}
\author{\bf Paul H. Frampton, Bettina Keszthelyi and Brian D. Wright}
\address{Institute of Field Physics, Department of Physics and Astronomy,\\
University of North Carolina, Chapel Hill, NC  27599-3255}
\author{\bf Thomas W. Kephart}
\address{Department of Physics and Astronomy, Vanderbilt University,
Nashville, TN 37235}
\maketitle
%\date{\today}

\begin{abstract}
By including the effects of superstring thresholds, we reconsider minimal
string unification
together with the requirement of producing
a supersymmetry-breaking gluino condensate in the hidden sector. This
gives, for examples of phenomenologically-viable $Z_8^{'}$ and other
related orbifolds, the constraint that
$0.1215 \leq \alpha_s(M_Z) \leq 0.1270$. In such models, a hidden photino
can be a source of cosmological dark matter detectable by gravitational
microlensing.

\end{abstract}

\newpage

Superstrings provide a hope for a mathematical framework
underpinning both particle theory and quantum gravity, yet
a testable prediction remains elusive.
Predictions concerning quantum gravity are generally too small
to be detectable but, of course, calculation of any parameter
in the standard particle theory would be an adequate confirmation.
We can, however, envisage a different empirical confrontation where
the hidden sector of the superstring impacts on the visible sector
and a connecting bridge at $M_{string}$ could then provide consistency checks.

Most attempts at superstring phenomenology over the last ten years have been
very speculative and, in any case, have concentrated on the link to
grand unification in the visible sector. By this we mean that in the heterotic
${\rm E}_8 \times {\rm E}_8^{'}$ superstring only, say, the first
${\rm E}_8$ is involved.
In this letter we use recent results for superstring thresholds to link
the visible sector to the hidden sector (${\rm E}_8^{'}$) and discuss the
implications of hidden sector supersymmetry breaking \cite{MR}\ for the value
of the
QCD coupling constant $\alpha_s(M_Z)$ and for cosmology. On the hidden
side we necessarily follow a top-down approach
because nothing is known empirically. The treatment of the
visible sector is partially bottom-up since we input
the low energy data.

In the visible sector, the standard model of particle theory is well
established up to
$100$ GeV and a supersymmetric grand unification at $M_{GUT} \sim 2 \times
10^{16}$ GeV,
corresponding to a supersymmetry breaking scale $\sim 1$ TeV, has been
advocated\cite{ADF,LL,EKN} (see, however, Ref. \cite{ADFFL}).
In a superstring the relevant scale is denoted $M_{string}$
which is related to Newton's gravitational constant ($M_{Planck}$ or the
reduced form
$\overline{M}_{Planck} = M_{Planck}/\sqrt{8\pi}$) through the string coupling
constant
$g_{string}$ in a calculable way. Here we shall assume an orbifold
compactification with, below $M_{string}$,
only the minimal supersymmetric standard model (MSSM) fields having standard
${\rm U}(1)_Y$ normalization. Since $M_{string}$ lies somewhat higher
(typically $M_{string}$ is a few times $10^{17}$ GeV) than $M_{GUT}$, a
successful minimal string unification at the higher scale necessitates
sizeable superstring threshold corrections.

The hidden sector is often assumed to generate supersymmetry breaking via
a gluino condensate for $\alpha_{hidden}(\mu_0) \rightarrow \infty$ at $\mu_0
\sim 10^{13}$
GeV\cite{condense}. At the same time this may generate Cold Dark Matter (CDM)
and have consequences for inflation and structure formation
in the visible sector. Big Bang Nucleosynthesis (BBN) constrains the massless
degrees of freedom in the hidden sector and thereby, in principle, the possible
breakings of the hidden ${\rm E}_8^{'}$; however, this will depend on the
ratio ($r$) of the temperatures of the hidden to visible sectors.

In the heterotic superstring $M_{string}$ is given by \cite{Kap,KL}
\begin{equation}
M_{string} = g_{string}\times 5.3 \times 10^{17} {\rm\ GeV}~,
\end{equation}
where $g_{string}$ is a dimensionless coupling constant related to the real
part of the
dilaton superfield expectation value by $g_{string}^{-2} = Re S$. Although,
according to
general arguments\cite{BD}, one expects the superstring
to be strongly coupled, one may also have the low energy string coupling
satisfying
$\alpha_{string} = g_{string}^2/4\pi < 1$ and hence a perturbative low energy
theory.
We shall see numerically that to obtain a reasonable supersymmetry breaking
scale,
$g_{string}\agt 0.2$, so $M_{string}$ must lie above the old GUT scale
(and becomes the new, higher, effective GUT scale) while lying safely below
$\overline{M}_{Planck}$.

The renormalization group equations for the ${\rm SU}(3)\times{\rm
SU}(2)\times{\rm U}_Y(1)$
gauge couplings $\alpha_a$ ($a = 1,2,3$) in the visible sector of the
superstring
have the form for $\mu \leq M_{string}$:
\begin{equation}
\frac{1}{\alpha_a(\mu)} =  \frac{k_a}{\alpha_{string}} +
\frac{b_a}{4\pi}\ln\frac{M_{string}^2}{\mu^2} + \frac{1}{4\pi}\Delta_a~,
\label{rge}
\end{equation}
where $k_1=\frac{5}{3}, k_2=k_3=1$.
Here $b_a$ are the one-loop renormalization group $\beta$ function coefficients
for the gauge groups and $\Delta_a$ represent the corresponding superstring
thresholds \cite{dkl,strthr}. In orbifold models,
$\Delta_a = - \sum_i b_a^{'(i)}\Delta^{(i)}$ where $\Delta^{(i)}$ are
threshold factors depending on the (1,1) string moduli for
the orbifold subplanes $i=1,2,3$
and $b^{'(i)}_a$ are related to the $\beta$ function
coefficients of the N = 2 sector of the orbifold. The latter are given
by\cite{ILR,IL}
\begin{equation}
b_a^{'(i)} = - C(G_a) + \sum_{R_a} T(R_a)(1+2n_{R_a}^{(i)})~,
\end{equation}
while, summing over the three subplanes gives
\begin{equation}
b_a^{'} = - 3C(G_a) + \sum_{R_a} T(R_a)(3+2n_{R_a})~,
\end{equation}
in which $T(R_a)$ are the Dynkin indices and $n_{R_a}$ are the modular weights
of the light matter fields in irreducible representations $R_a$.
If the (1,1) moduli $T^i$ are {\it either} all equal
$T = T^1 = T^2 = T^3 $ (isotropic orbifold) {\it or} one of them is by far the
largest $T = T^1 \gg T^2, T^3$ (squeezed orbifold)
then we may write $\Delta_a = b_a^{'}\Delta$ where $b_a^{'} = \sum_i
b_a^{'(i)}$
for the former case and $b_a^{'} = b_a^{'(1)}$ for the latter case.
In either case, $\Delta$ is a common function whose explicit form in the large
$T$
limit is given for simple cases by\cite{dkl}
\begin{equation}
\Delta = \ln (|T + \overline{T}||\eta(T)|^4)
\label{orbithr}
\end{equation}
where $\eta$ is the Dedekind function.

Elimination of $\alpha_{string}$ in Eqs.~(\ref{rge}) gives the generalization
of
the GQW equations\cite{GQW}:
\begin{equation}
\sin^2\theta_W(M_Z) = \frac{k_2}{k_1 + k_2} - \frac{k_1}{k_1 + k_2}
\frac{\alpha_{em}(M_Z)}{4\pi}\left[A\ln\left(\frac{M_{string}^2}{M_Z^2}\right)
- A^{'}\Delta\right],
\label{sw}
\end{equation}
\begin{equation}
\alpha^{-1}_s(M_Z) = \frac{k_3}{k_1+k_2}\left[\alpha^{-1}_{em}(M_Z) -
\frac{1}{4\pi}B\ln\left(\frac{M^2_{string}}{M_Z^2}\right) +
\frac{1}{4\pi}B^{'}\Delta\right],
\label{as}
\end{equation}
where $A = \frac{k_2}{k_1}b_1 - b_2$, $B = b_1 + b_2 - \frac{k_1+k_2}{k_3}b_3$
and $A^{'},B^{'}$
are obtained from $A,B$ by the substitution $b_a \rightarrow b_a^{'}$.
For the MSSM one has the values $A=\frac{28}{5}, B = 20$.

The renormalization group equation for the coupling constant in the hidden
sector
corresponding to any subgroup $G^{'}$ of ${\rm E}_8^{'}$ is
\begin{equation}
{1\over\alpha_{G^{'}}(\mu)} = {1\over\alpha_{string}} +
\frac{b_{G^{'}}}{4\pi}\ln\frac{M_{string}^2}{\mu^2}
\label{hiddenrge}
\end{equation}
Here we have assumed no matter fields are present in the hidden sector,
so the corresponding superstring thresholds vanish.
The value of $\mu = \mu_0$ where $\alpha_{G^{'}}(\mu_0) \rightarrow \infty$ is
thence given by:
\begin{equation}
\mu_0 = M_{string}\exp\left[\frac{2\pi}{\alpha_{string}b_{G^{'}}}\right].
\label{muzero}
\end{equation}
The scale $\mu_0$ is determined by assuming the gravitino mass arises from the
standard
effective supergravity coupling between the gravitino and the
gaugino bilinear ($\langle\chi\chi\rangle \sim \mu_0^3$).
This gives a gravitino mass (within an order of magnitude)
\begin{equation}
m_{3/2} \sim {\mu_0^3/\overline{M}_{Planck}^2} = 100 {\rm\ GeV\ -\ } 1{\rm\
TeV}~,
\label{mgrav}
\end{equation}
to obtain the correct visible sector supersymmetry breaking. We will consider
values
in the range $\mu_0 =  (0.4 {\rm -} 4) \times 10^{13}$ GeV taking into account
the above
range and order of magnitude factors. To obtain the hidden sector gaugino
condensate at
{\it e.g.} $\mu_0 \sim 10^{13}$ GeV
(corresponding to visible supersymmetry breaking at $\sim 1$ TeV) we consider
the quadratic Casimir, $C_2(G^{'}) = -\frac{1}{3}b_{G^{'}}$, ranging from 2
(for SU(2))
to 30 (for E(8)). Recall that $C_2(G^{'})$ = N for SU(N)
and 12,18,30 for ${\rm E}_{6,7,8}$ respectively.
The results of solving Eq.~(\ref{muzero}) for $M_{string}$ are displayed in
Fig.~1
for the above range of $\mu_0$.

Given $M_{string}$ from the hidden sector analysis and a value
of $\gamma = B^{'}/A^{'}$, we can now determine from Eqs.~(\ref{sw},\ref{as})
the value
of $A^{'}\Delta$ and $\alpha_s(M_Z)$ respectively,
for the range of allowed empirical values for $\sin^2\theta_W(M_Z)$ and
$\alpha_{em}(M_Z)$ which, taking into account low energy MSSM, are
currently\cite{L}
\begin{eqnarray}
\sin^2\theta_W(M_Z) & = & 0.2313\pm0.0003 \nonumber\\
\alpha_{em}^{-1}(M_Z) & = & 128.09\pm0.09~.\label{ewparam}
\end{eqnarray}
Since Eqs.~(\ref{sw},\ref{as}) determine $\alpha_s(M_Z)$ only in terms of the
ratio
$\gamma$, we must try to determine $\gamma$. As a first example we use the
compactification of the heterotic string on the $Z_8^{'}$ orbifold,
as discussed in\cite{ILR,IL}. In that particular example, for the three complex
planes
corresponding to the three two-dimensional subtori within the orbifold, the
MSSM states
were assigned modular weights as follows: $n_{Q_{1,2,3}}=(0,-1,0);
n_{D_{1,2,3}}=(-1,0,0);
n_{U_1}=(0,-\frac{1}{2},-\frac{1}{2}); n_{U_{2,3}}=(-\frac{3}{4},
-\frac{15}{8},-\frac{3}{8}); n_{L_1}=(-\frac{14}{8},
-\frac{3}{8},-\frac{7}{8});
n_{L_{2,3}}=(-\frac{14}{8}, -\frac{7}{8},-\frac{3}{8}); n_{E_1}=(-1,0,0);
n_{E_{2,3}}=
(-\frac{3}{4}, -\frac{15}{8},-\frac{3}{8}); n_H=(-\frac{1}{2},
-\frac{3}{4},-\frac{3}{4});
n_{\overline{H}}=(-\frac{14}{8}, -\frac{3}{8},-\frac{7}{8})$.
Using the definitions given above for the separate orbifold subplanes $i =
1,2,3$,
this gives $b_{1}^{'(1,2,3)}=-\frac{15}{2},-\frac{145}{2},\frac{35}{2}$
yielding $b_1^{'}=-\frac{50}{3}$;
$b_{2}^{'(1,2,3)}=-\frac{5}{2},-\frac{29}{4},\frac{7}{4}$ yielding
$b_2^{'}=-8$; and
finally $b_{3}^{'(1,2,3)}=-\frac{3}{2},-\frac{29}{4},\frac{7}{4}$ yielding
$b_3^{'}=-7$.
In this case, the values of $A^{'},B^{'}$ defined above are $A^{'} = -2$ and
$B^{'} = -6$
giving $\gamma = 3$.  For the other viable orbifolds cited in \cite{IL} the
results
are similar. For possible assignments of modular weights in squeezed $Z_6$
and $Z_2 \times Z_2$ orbifold models one can find $\gamma = 3$, just as in the
$Z_8^{'}$ case.

Of course, a range of values for $\gamma$ is possible for a specific
construction. But for all these
special choices of modular weights, one has $\gamma = 3$ so in Fig.~2 we
show the relationship between $C_2(G^{'})$ and $\alpha_s(M_Z)$ for this value
of $\gamma$ and
taking $\mu_0=10^{13}$ GeV.  Curves are shown for the central
values of $\alpha_{em}(M_Z)$ and $\sin^2\theta_W(M_Z)$ and for the extremes of
the
allowed ranges. Note that $\alpha_s(M_Z)$ is maximized for $\alpha_{em}(M_Z)$
maximized and $\sin^2\theta_W(M_Z)$ minimized, and
{\it vice versa}. The shifted short(long) dashed curves in Fig.~2
correspond to $\mu_0 = 0.4(4.0)\times 10^{13}$ GeV,
showing that the results are relatively insensitive to $\mu_0$.
The range of $\alpha_s(M_Z)$, if we assume $C_2({\rm SU}(2))=2 \leq C_2(G^{'})
\leq C_2({\rm E}_7)=18$, to allow the CDM hidden photino discussed below,
is seen from Fig.~2 (for $\mu_0 = 10^{13}$ GeV) to be
\[ 0.1215 \leq \alpha_s(M_Z) \leq 0.1270~. \]
This is to be compared to the allowed LEP range $\alpha_s(M_Z) = 0.121 \pm
0.005$\cite{L}.

In the above, we have taken the particular value $\gamma = 3$ which occurs in
three simple
examples given in \cite{IL} seriously. If we instead regard $\gamma$ as an
arbitrary
rational number and input $\mu_0 = 10^{13}$ GeV, the full allowed ranges
of $\alpha_{em}(M_Z), \sin^2\theta_W(M_Z)$ and $\alpha_s(M_Z) = 0.121 \pm
0.005$ \cite{L}
then we find $2.62(2.5) \leq \gamma \leq 3.73(3.75)$ for $2 \leq C_{2}(G^{'})
\leq 18(30)$.
This is illustrated in Fig.~3 where the
three cases $\gamma = 2.62, 3.00$ and $3.73$ are shown with uncertainties as in
Fig.~2.
Thus Fig.~3 provides constraints on $\gamma$
and $C_{2}(G^{'})$ for orbifold constructions to be consistent with LEP data,
the gaugino
condensate idea and the hidden photino CDM candidate disussed below.
In any specific orbifold compactification, the quantities
$\gamma$ and $C_2(G^{'})$ may be determined (or chosen from a range consistent
with
modular invariance) and therefore $\alpha_s(M_Z)$ is fixed up to a narrow range
coming from uncertainties in (\ref{ewparam}) and an allowed range of $\mu_0$.
In principle, one should do a full two loop analysis and include the effects of
a detailed MSSM spectrum in the form of low energy threshold corrections.
However,
given the uncertainties at $M_{string}$, these are less important
in this type of analysis.

All these results depend on being able to obtain the necessary value for the
moduli-dependent superstring threshold correction $\Delta$ determined by
Eqs.~(\ref{sw},\ref{as})
in a given orbifold construction. For the combination $A^{'}\Delta$, the
solution
is mostly sensitive to $C_2(G^{'})$, or equivalently $M_{string}$. For example,
for
$2 \leq C_2(G^{'}) \leq 18$ and $\mu_0 = 10^{13}$ GeV we find
$38 \geq A^{'}\Delta \geq 25$, independently
of $\gamma$.  Given the value of $A^{'}$, there may be
naturalness constraints on the size of $|\Delta|$ coming from the values of the
moduli
VEVs by which it is determined, so one might be restricted to large hidden
sector gauge groups.
However, as shown in \cite{NS}, it is possible to obtain sufficiently
large $|\Delta|$ in a $Z_8^{'}$ orbifold with $A^{'} = -2$ for natural values
of the moduli
VEVs by including continuous Wilson lines; the latter are, in any case,
generally necessary for the
required symmetry breaking pattern in both the visible and hidden sectors.

Now we turn to the cosmological aspects and ramifications for
the hidden sector. From the above, an illustrative scenario is
where $g_{string} \sim 0.7$ and the condensate occurs
in an SU(5) gauged subgroup of ${\rm E}_8^{'}$; for example, the breaking
of ${\rm E}_8^{'}$ by Wilson lines could give the rank 8 subgroup ${\rm SU}(5)
\times
{\rm SU}(4) \times {\rm U}(1)$. The SU(4) condensates are sufficiently heavy
to decay gravitationally before BBN\cite{KN} and the hidden photino associated
with
the U(1) will have a mass comparable to the visible supersymmetry breaking
scale of
$\sim 1$ TeV. Let us now pursue the general idea that such a
shadow photino is the origin of some, or all, cosmological dark matter.

A serious constraint on the number of hidden sector massless degrees of freedom
arises from the agreement of the visible standard model with BBN.
The hidden sector photons change the effective number of degrees  of freedom
according to:
\begin{equation}
g_{eff}^* = g_{visible}^* +
g_{hidden}^*\left(\frac{T_{hidden}}{T_{\gamma}}\right)^4
\label{geff}
\end {equation}
where $g_{visible}^*$ is the visible sector degrees of freedom at
nucleosynthesis
(10.75 for $e^{\pm}$(3.5), $\nu_i$(5.25) and $\gamma$(2.0))
while $g_{hidden}^* = 2N_{{\gamma},hidden}$ for $N_{{\gamma},hidden}$ hidden
photons.
The upper limit on $g_{eff}^*$ is $\sim 11.45$\cite{CST} and so there cannot be
any shadow photon unless $T_{hidden} < T_{\gamma}$; if $T_{hidden} =
T_{\gamma}$,
no such extra massless state is permissible. (Note that in the older
literature\cite{KST}
the upper limit on $g_{eff}^*$ was weaker: $g_{eff}^* < 13$).

As far as Eq.~(\ref{geff}) and BBN are concerned, a key issue is the ratio of
hidden to visible temperatures at the BBN era.
If we adopt a hidden photino mass in the range 100 GeV - 1 TeV, and assume
that it forms all the dark matter ($\Omega_{\tilde{\gamma}} \sim 1$), then the
temperature ratio
$(T_{hidden}/T_{\gamma}) \sim  10^{-3.3}$ to $10^{-3.7}$, since
the number density of hidden photinos $n_{\tilde{\gamma},hidden} \propto
T_{hidden}^3$
just as $n_{\gamma} \propto T_{\gamma}^3$ and we assume
$n_{\tilde{\gamma},hidden} =
\frac{3}{4}n_{\gamma}$ at a very early era when $T = T_{\gamma} = T_{hidden}$
close to $T = M_{Planck}$.

Consider a model with normal sector gauge group $G$ and
hidden sector gauge group $G_{hidden}$, whose origins are both
in our heterotic superstring theory.  The symmetry breaking patterns, and
herefore the phase transition structure of such a model, can be
quite complicated, and the temperatures $T_{\gamma}$ and $T_{hidden}$ of the
two sectors can evolve very differently, but are typically
intertwined by inflation.

As an example, consider a scenario in which there are two inflationary epochs.
Suppose that an inflaton, {\it e.g.} the dilaton, induces a period of inflation
affecting equally the visible and hidden sectors.
The universe supercools exponentially since the scale factor increases
exponentially and $RT$ = constant. Most
of the required e-foldings may be accomplished during this inflationary phase.
The final, say, $\sim 8$ e-foldings may arise from an inflation induced,
for example, by a superstring modulus field which couples to the visible matter
and not to the hidden sector. This second inflationary phase can
occur at the weak scale as suggested by Ref.~\cite{RT} to solve the
cosmological moduli problem.
Reheating can then occur in the visible sector
back to approximately the critical temperature, while no reheating is
possible in the hidden sector.
This leads to a temperature ratio between hidden and visible sectors $e^{-8}
\sim 10^{-3.5}$ as required. This is just one possible scenario to illustrate
how
enough dark matter could be generated to make $\Omega = 1$.

Finally, we come to the testability of our dark matter proposal.
The hidden photinos have a Jeans mass, $M_{Jeans}$, given by $M_{Jeans}=
M_{Planck}^3/(m_{\tilde{\gamma},hidden})^2 \sim 10^{51}$ GeV $\sim
10^{-6}M_{\odot}$
for $m_{\tilde{\gamma},hidden} = 1$ TeV;
for the lower value $m_{\tilde{\gamma},hidden} = 100$ GeV,
$M_{Jeans} \sim 10^{-4}M_{\odot}$.
Since this is CDM we expect it to have clumped gravitationally
at this scale and so our galactic halo will be comprised of these hidden
objects.
The accretion of hidden photinos into such MACHOs will cause gravitational
microlensing of distant stars, and  detection
of the resultant temporary achromatic amplification is a possibility.
The method\cite{P,G} is practical for dark matter objects in the
range $10^{-7}M_{\odot}$ to $10^2M_{\odot}$.

The duration of a microlensing event scales as $M^{\frac{1}{2}}$, and
since for $\sim 10^{-1}M_{\odot}$, the observed event durations
are a few weeks, the shadow photino MACHOs should have microlensing event
durations of order a few hours or days. Detection of such events by
dedicated searches would support the interpretation suggested here
as the origin of MACHOs in the required mass range. Although the details are
model
dependent, such ${\rm U}(1)$ factors in the hidden sector are generic in
superstring theories and hence give rise to such CDM candidates.

This work was supported in part by the U.S. Department of
Energy under Grants DE-FG05-85ER-40219, Task B, and
DE-FG05-85ER-40226.

\begin{figure}
\caption{$M_{string}$ as a function of the hidden sector gauge group
subject to the gaugino condensation condition. Solid lines correspond to
$\mu_0 = 10^{13}$ GeV while short(long) dashed lines are for
$\mu_0 = 0.4(4.0)\times 10^{13}$ GeV.}
\label{fone}
\end{figure}

\begin{figure}
\caption{$\alpha_s(M_Z)$ as a function of the quadratic Casimir for
the
highest rank hidden sector gauge group for an orbifold model with $\gamma = 3$
and taking $\mu_0 = 10^{13}$ GeV (solid curves). The upper and lower solid
curves take
into account the uncertainties in $\sin^2\theta_W$ and $\alpha_{em}$.
The shifted short(long) dashed curves indicate the effect of
changing $\mu_0$ to $0.4(4.0)\times 10^{13}$ GeV.}
\label{ftwo}
\end{figure}

\begin{figure}
\caption{$\alpha_s(M_Z)$ as a function of the quadratic Casimir for
the highest rank hidden sector gauge group for three values of
$\gamma$ (2.6,3.0,3.7)
and taking $\mu_0 = 10^{13}$ GeV. The horizontal lines give the current LEP
limits
on $\alpha_s(M_Z)$ assuming the MSSM.
The range $2.62 < \gamma < 3.73$ is that which just allows at least one ${\rm
U}(1)$
component in the hidden sector gauge group. Uncertainties in $\sin^2\theta_W$
and
$\alpha_{em}$ are indicated by the shaded regions.}
\label{fthree}
\end{figure}

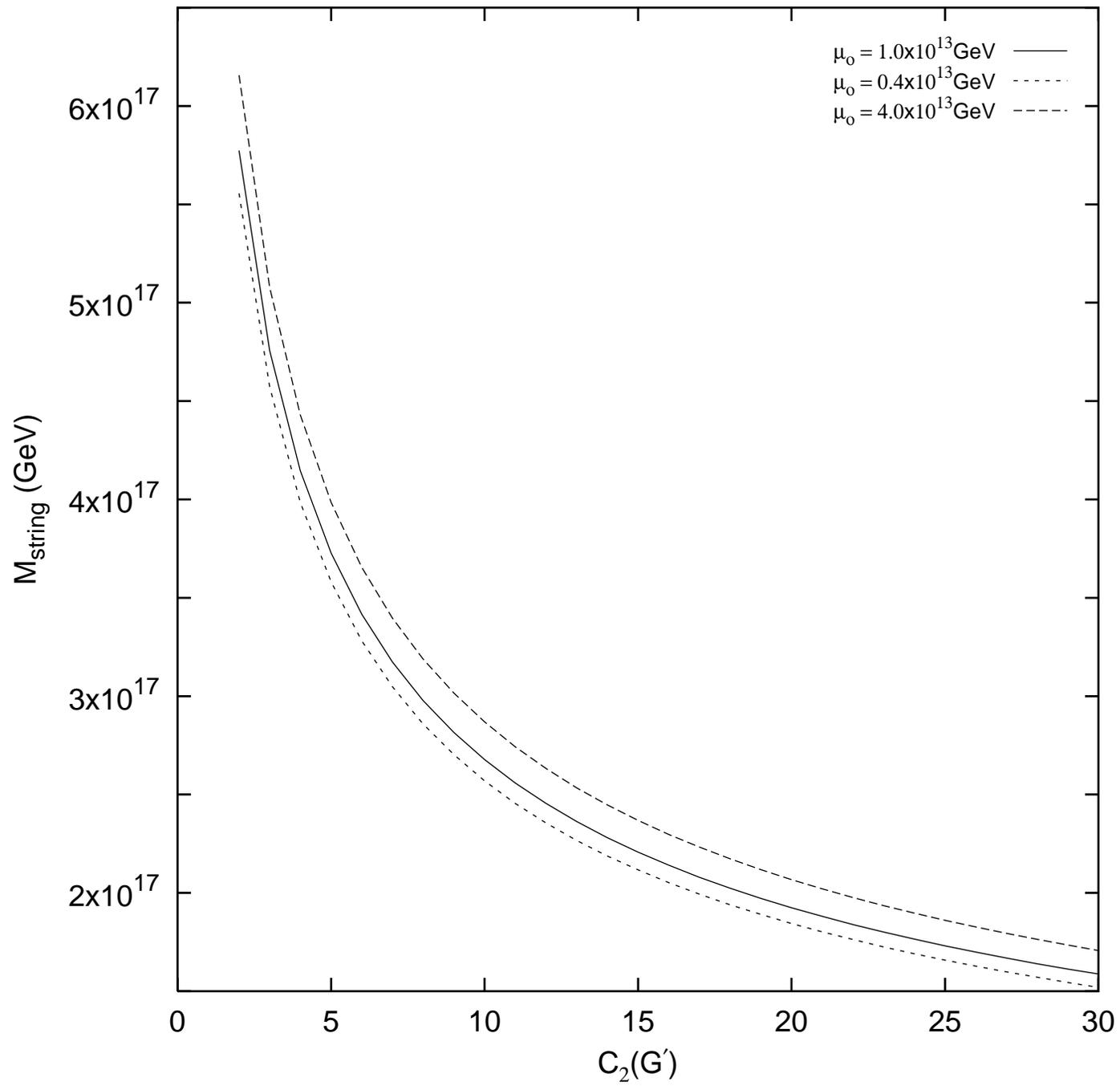
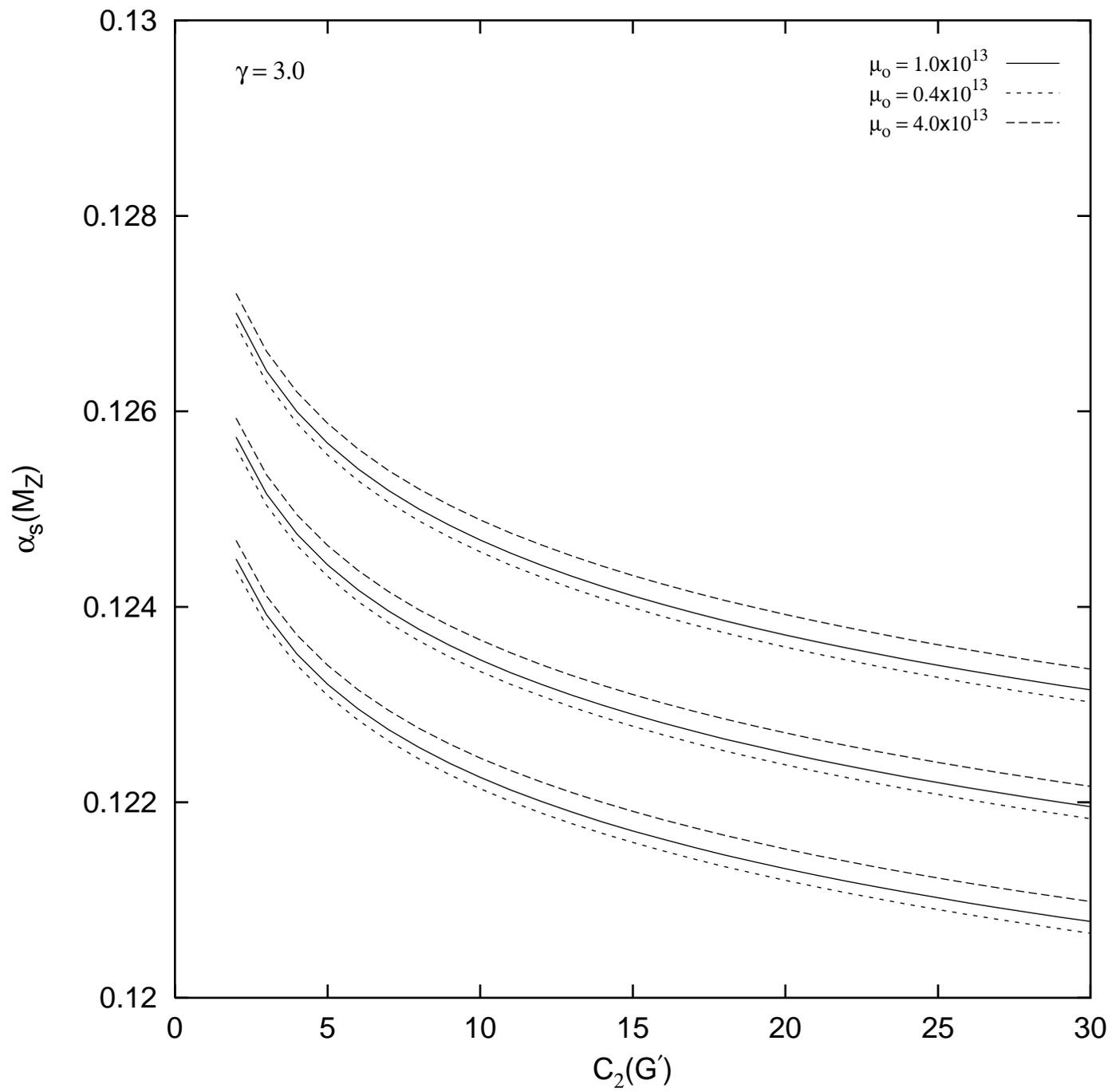
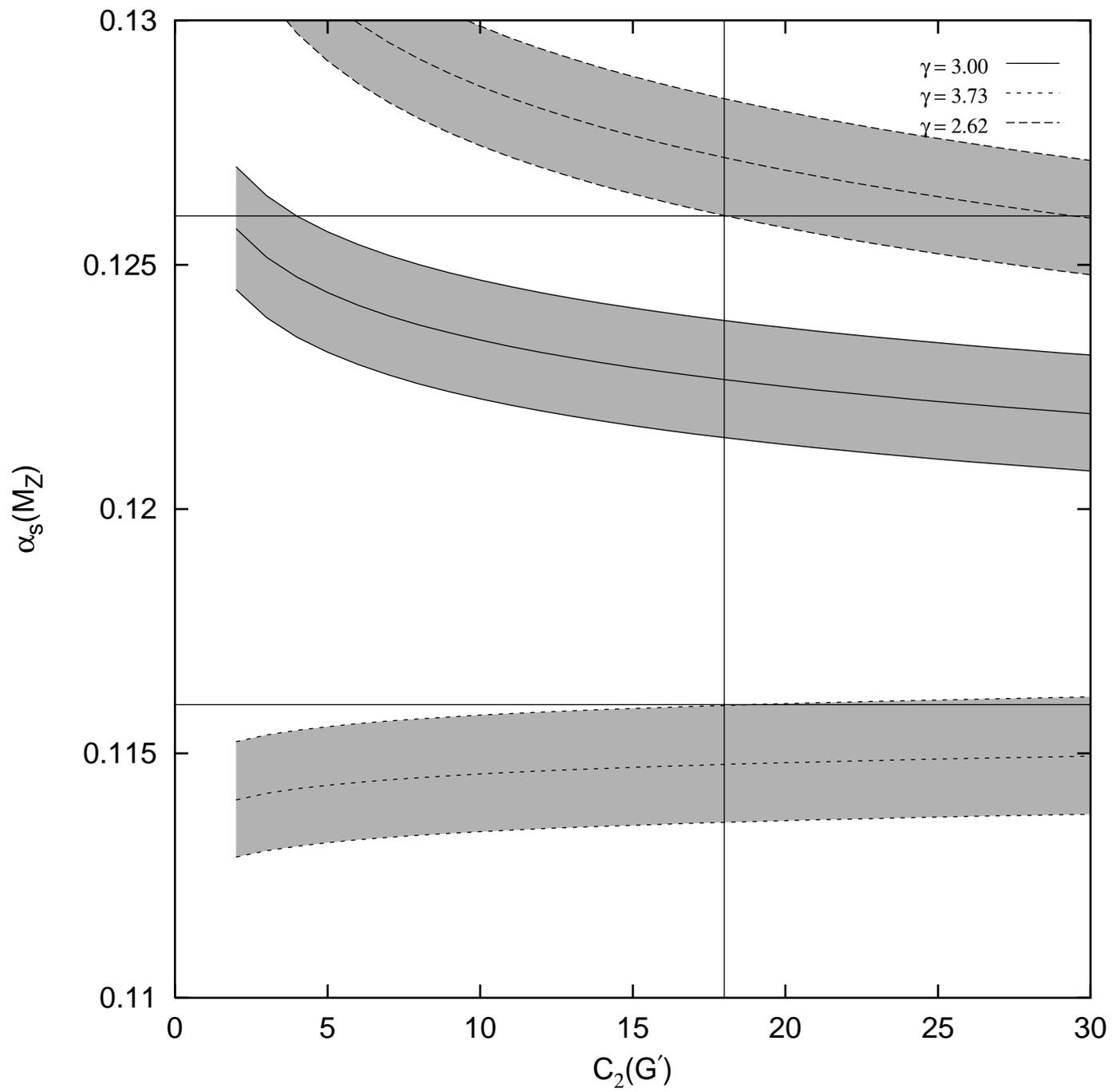
\end{document}